\documentclass[a4paper]{llncs}
\usepackage[top=2.05in, bottom=2.05in, left=1.73in, right=1.73in]{geometry}
\usepackage{amssymb}
\usepackage{graphicx}
\usepackage{color}

\usepackage{url}
\urldef{\mailsa}\path|{alfred.hofmann, ursula.barth, ingrid.haas, frank.holzwarth,|
\urldef{\mailsb}\path|anna.kramer, leonie.kunz, christine.reiss, nicole.sator,|
\urldef{\mailsc}\path|erika.siebert-cole, peter.strasser, lncs}@springer.com|    
\newcommand{\keywords}[1]{\par\addvspace\baselineskip
\noindent\keywordname\enspace\ignorespaces#1}

\usepackage{tabularx}
\newcolumntype{L}[1]{>{\raggedright\arraybackslash}p{#1}}
\newcolumntype{C}[1]{>{\centering\arraybackslash}p{#1}}
\newcolumntype{R}[1]{>{\raggedleft\arraybackslash}p{#1}}

\usepackage{booktabs}
\usepackage{cite}
\usepackage{subfigure}
\usepackage{amsmath}
\usepackage{graphicx}
\begin{document}

\mainmatter  

\title{Application of Top-hat Transformation for Enhanced Blood Vessel Extraction}

%
%
\author{Tithi Parna Das$^1$ \and Sheetal Praharaj$^1$ \and Sarita Swain$^1$ \and Sumanshu Agarwal$^1$ \and Kundan Kumar$^2$\thanks{corresponding author}%
}
%


\institute{Department of Electronics and Communication Engineering, ITER, Siksha `O' Anusandhan (Deemed to be University), Bhubaneswar-751030, Odisha, India
\and Center for the Internet of Things, ITER, Siksha `O' Anusandhan (Deemed to be University), Bhubaneswar-751030, Odisha, India\\\email{tithiparna100@gmail.com, sheetalpraharaj99@gmail.com, sarita826011@gmail.com,sumanshuagarwal@soa.ac.in, kundankumar@soa.ac.in}\\\url{https://erkundanec.github.io/} 
}

%
%

\tocauthor{Authors' Instructions}
\maketitle

\begin{abstract}
In the medical domain, different computer-aided diagnosis systems have been proposed to extract blood vessels from retinal fundus images for the clinical treatment of vascular diseases. Accurate extraction of blood vessels from the fundus images using a computer-generated method can help the clinician to produce timely and accurate reports for the patient suffering from these diseases. In this article, we integrate top-hat based preprocessing approach with fine-tuned B-COSFIRE filter to achieve more accurate segregation of blood vessel pixels from the background. Use of top-hat transformation in the preprocessing stage enhances the efficacy of the algorithm to extract blood vessels in presence of structures like fovea, exudates, haemorrhages, etc. Furthermore, to reduce the false positives, small clusters of blood vessel pixels are removed in the postprocessing stage. Further, we find that the proposed algorithm is more efficient as compared to various modern algorithms reported in the literature.
\keywords{Top-hat transform, B-COSFIRE filter, Blood vasculature, Retinal vessel segmentation, Retinal blood vessel extraction.}
\end{abstract}

\section{Introduction}
\label{Intro}
\begin{figure}
     \centering
     \includegraphics[width=0.5\textwidth]{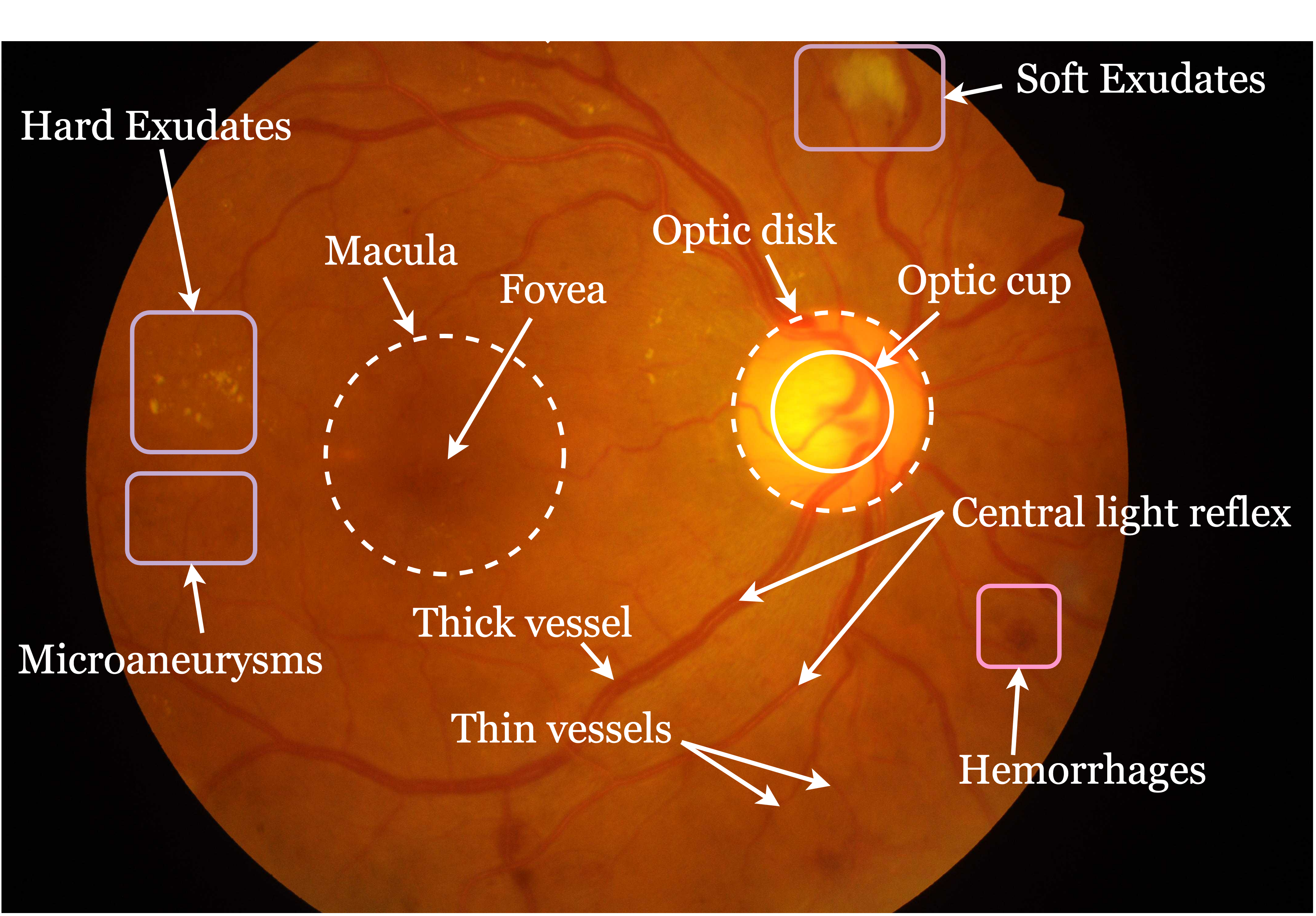}
     \caption{Labeled diagram of Retinal fundus image}
     \label{fig:labeledDiagram}
 \end{figure}

 Retinal fundus photography facilitates the examination of several vascular diseases, such as age-related macular degeneration (AMD), diabetic retinopathy (DR), glaucoma, arteriosclerosis, hypertension, etc., through non-invasive procedure~\cite{Chaudhuri1989,mookiah2021a}. These diseases can be monitored periodically by tracking the change in thickness of blood vessels and/or new branches in the vasculature of retinal blood vessels (a tree-like structure)~\cite{Soares2006}. However, the extraction of vasculature from an irregularly illumined background in presence of exudates, haemorrhages, optic disc, etc. is a challenging task. Moreover, the central reflex in retinal blood vessels makes the extraction process more difficult~\cite{Fraz2012} (see Figure~\ref{fig:labeledDiagram}).
 
 \par Although several supervised and unsupervised blood vessel extraction procedures have been presented in the literature, most of them are not capable to address all the challenges together~\cite{Hoover2000, Zhang2010, 8989144}. Supervised approaches use the pixel class levelled images (ground truth) to design a model to distinguish between a blood vessel and background pixels, while unsupervised approaches entirely confide on the pixel values of the input images to design a blood vessel extraction model. Artificial Neural Network (ANN), Support Vector Machine (SVM), k-NN-classifier are some of the well established supervised learning models that have been trained utilizing the image response of the line detector, grey level co-occurrence matrix (GLCM), and Gabor filters. However, in state-of-art algorithms, extraction of these handcrafted features have been replaced with an automated filtering process through Convolutional Neural Network (CNN) model and then different classification models were used to categorize blood vessel and background pixels~\cite{Fu2016a, Jin2019a, Li2020}. Although these models offered state-of-art findings; these methods are complex and time-consuming than demand a large number of images to train the model~\cite{mookiah2021a}.
 
\par On the other hand, an unsupervised method based on filtering approaches received great attention in the domain such as 2-D Gaussian kernel, matched filters, the first-order derivative of Gaussian, B-COSFIRE filters, etc. These approaches rely on the hypothesis that the cross-sectional pixel intensity of a retinal blood vessel has the Gaussian/rectangular-shaped profile. Chaudhuri \textit{et al.} have applied a 2-D Gaussian kernel for segregate blood vessels from the background~\cite{Chaudhuri1989}. Based on the concept of matched filtering, Hoover \textit{et al.} have designed a piecewise threshold probing algorithm considering some region-based properties~\cite{Hoover2000}. Zhang \textit{et al.} have proposed a combination of a first-order derivative of the Gaussian filter and zero-mean Gaussian which resolves some of the limitations of previous methods~\cite{Zhang2010}. Another variation proposed by Odstrcilik \textit{et al.} where illumination correction and contrast equalization were performed before matched filtering~\cite{Odstrcilik2013}. Azzopardi \textit{et al.} have proposed the B-COSFIRE filtering approach, which is better than many lineation algorithms under the unsupervised category~\cite{Azzopardi2015}. In spite of that, the approach is not very efficient for pathological fundus images. In most of the algorithms, whether it belongs to a supervised or unsupervised category, the input images are preprocessed through contrast limited histogram equalization (CLAHE) technique to enhance the contrast between blood vessel pixels and the background. Recently, Kumar \textit{et al.} have proposed a top-hat morphological operation based preprocessing technique that suppress the other structures present in the background~\cite{Kumar2019}.
\par Here, we integrate the top-hat based preprocessing method with fine-tuned B-COSFIRE filter to eliminate the other structure present in the background especially in the case of pathological retinal images. During the preprocessing, the proposed approach uses the top-hat morphological operation followed by the CLAHE enhancement technique for better contrast between the blood vessel and background pixels. This advanced preprocessing approach is very effective when other structures like exudates, optic disc, haemorrhages are present in the background. Furthermore, the B-COSFIRE filter is fine-tuned accordingly for best performance. The experimental evaluation and comparative analysis are performed to show the superiority of the proposed approach for retinal blood vessel extraction.

Organisation of the paper is as follows. The different steps involved for retinal blood vessel extraction are discussed in section 2. The section 3 discusses the dataset used and experimental outcomes. The conclusion is discussed thereon.

\section{Proposed method}
\label{Prop}
The proposed approach is the integration of preprocessing step and B-COSFIRE filtering process followed by post-processing. The following sections briefly describe the steps involved to extract the vascular tree structure from retinal fundus images.

\subsection{Preprocessing}
\begin{figure}[htbp]
    \centering
    \subfigure[]{\includegraphics[scale=0.12]{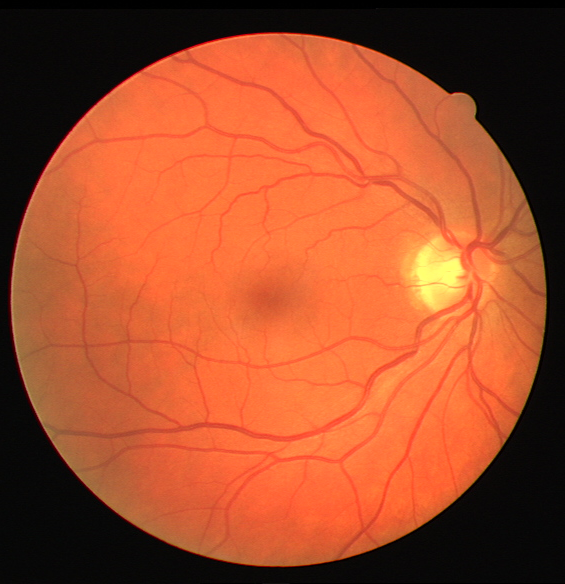}\label{fig_orig}}~~
    \subfigure[]{\includegraphics[scale=0.12]{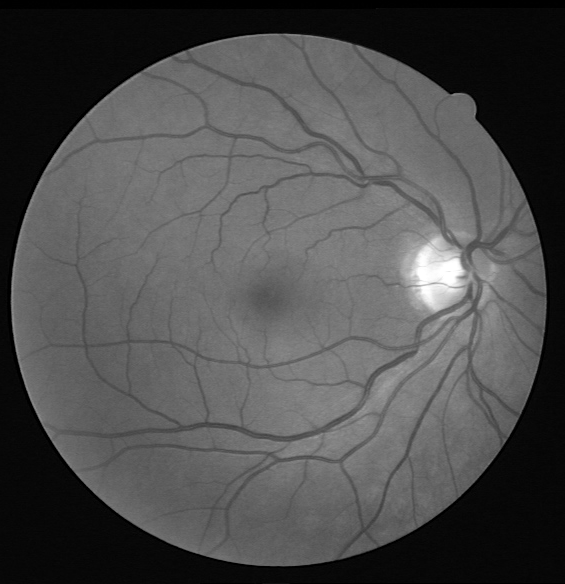}\label{fig_green}}~~
    \subfigure[]{\includegraphics[scale=0.12]{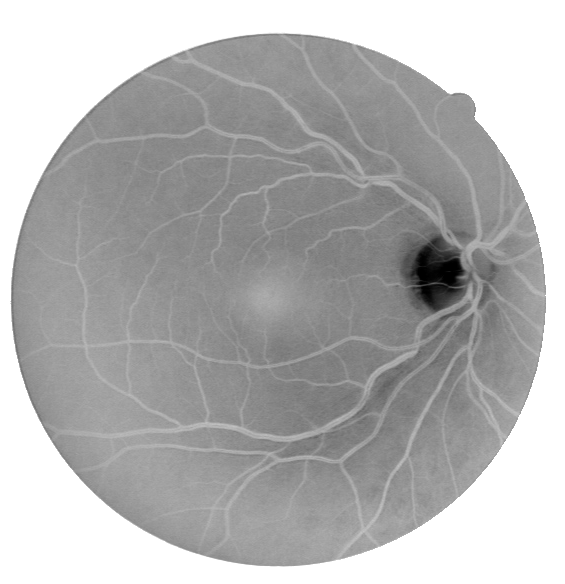}\label{fig_greeninv}}~~
     \subfigure[]{\includegraphics[scale=0.12]{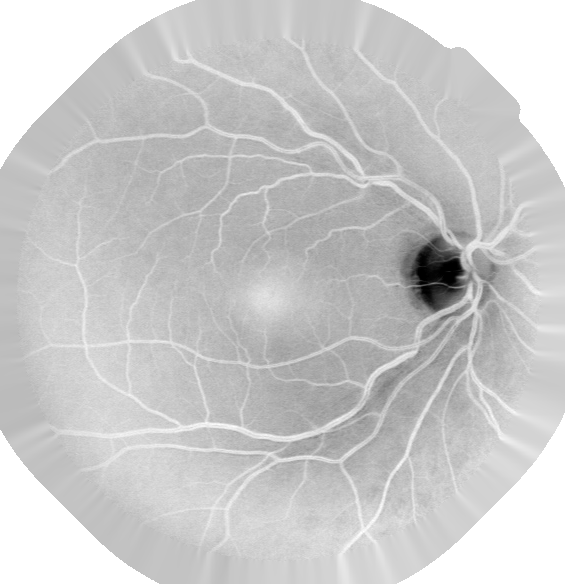}\label{fig_fakepadded}}\\
     \subfigure[]{\includegraphics[scale=0.12]{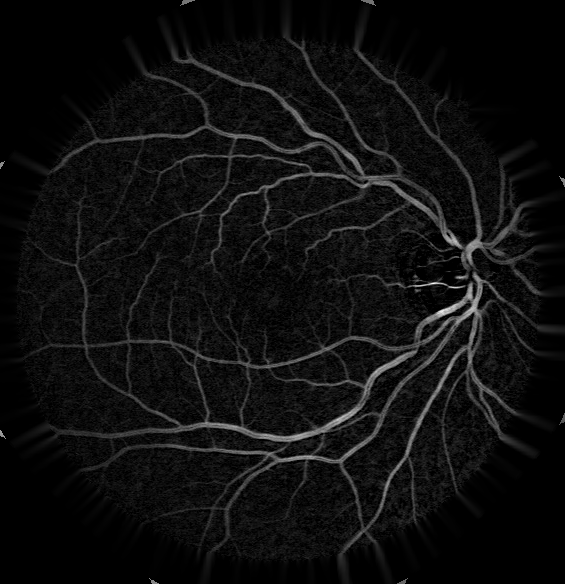}\label{fig_Tophat}}~~
      \subfigure[]{\includegraphics[scale=0.12]{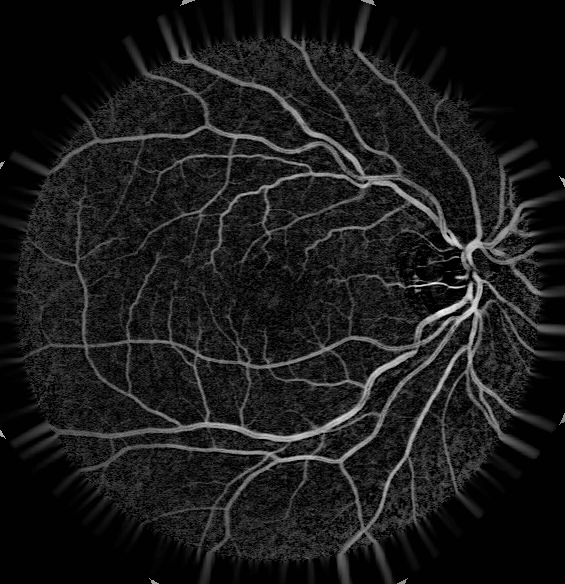}\label{fig_Clahe}}~~
       \subfigure[]{\includegraphics[scale=0.12]{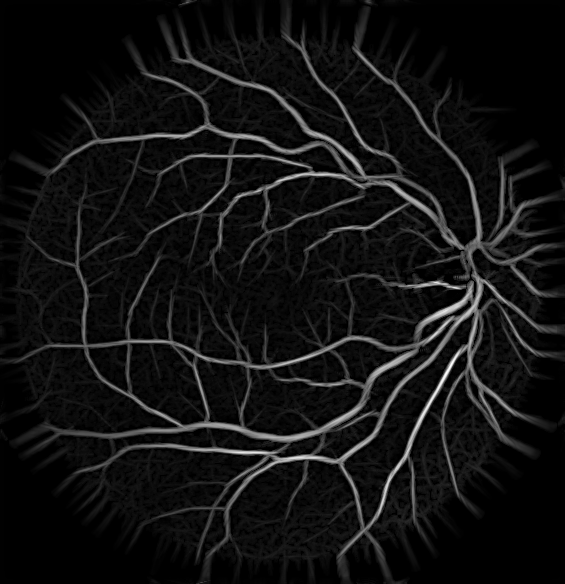}}~~
     \subfigure[]{\includegraphics[scale=0.12]{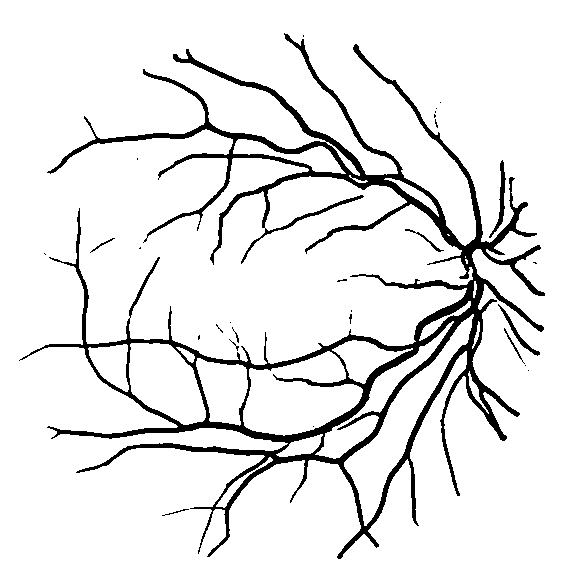}}
    \caption{Intermediate results after processing (a) input image, (b) green channel, (c) inverted green channel, (d) fake padded image, (e) image after top-hat transform, (f) CLAHE response, (g) feature response after applying B-COSFIRE filter, and (h) binary segmented image using Otsu's method.}
    \label{fig_preproc}
\end{figure}

For retinal fundus image processing, handling uneven illuminance, central light reflex diminution, border pixels, and presence of other structures along with blood vessels are major challenges as stated earlier. Different preprocessing techniques can be used to deal with some of the challenges. In this work, we preprocess the grayscale fundus image using two different image enhancement techniques. Before that green channel image is separated from the colour fundus image as the grayscale fundus image. In the green channel image, pixels from the blood vessels appear darker than background pixels. To make the blood vessel pixel bright, the grayscale image is inverted (see Figure~\ref{fig_greeninv}).

Next, we obtain a fake padded image ($I_G$) to radially extend the border pixel of the field-of-view (FOV) area as shown in Figure~\ref{fig_fakepadded}. This step enhances the rejection of false-positive detection of vessel pixels around the border. After fake padding, the image is processed through white-top-hat morphological transformation~\cite{Kumar2019} given as
\begin{equation}
    I_{{TH}} = I_G-I_G\circ e
\end{equation}
where $I_{TH}$ is the top-hat (TH) transformed image through structuring element $e$ and $\circ$ is the opening operator. 
Top-hat transformation followed by CLAHE further improved the contrast between blood vessel pixels and the background (see Fig.~\ref{fig_Clahe}). Before proceeding with the further processing of the data, we normalized pixel values of preprocessed images between 0 and 1.  

\subsection{B-COSFIRE Filtering}
\par The preprocessed retinal fundus image is further processed through B-COSFIRE (Bar-selective Combination Shifted Filter Responses) filters proposed by Azzopardi \textit{et al.}~\cite{Azzopardi2015}, which is based on basic COSFIRE filters~\cite{Azzopardi2013}. Because the blood vessels are bar-like structures and the same is detected in the images by the filter, therefore, it is called as bar-selective. The response of the B-COSFIRE filters are obtained by the weighted geometric mean of responses of difference of Gaussian (DoG) filters with different position configurations. The positions configuration, at which the DoG responses are calculated, are fully automatic. The response of the B-COSFIRE filter is obtained by performing the following four steps.
\begin{itemize}
    \item[(i)] \textit{Detection of Intensity Change}:
A center-on DoG function, denoted as $\text{DoG}_\sigma(x,y)$ is obtained as
    \begin{align}
        \text{DoG}_{\sigma}(x, y) \stackrel{\text { def }}{=}& \frac{1}{2 \pi  \sigma^2} \exp \left[-\frac{\left(x^{2}+y^{2}\right)}{2 \sigma^2}\right]\nonumber\\
        &-\frac{1}{2 \pi\left(0.5 \sigma)^2\right.} \exp \left[-\frac{\left(x^{2}+y^{2}\right)}{2\left(0.5 \sigma\right)^2}\right]
    \end{align}

where $\sigma$ is a hyper-parameter that is to be tuned to decide an excitatory central region and inhibitory surround. The  DoG filter response is obtained by convolving the image $I$ with a kernel function $\text{DoG}_\sigma(x-x',y-y')$ as

\begin{equation}
c_{\sigma}(x, y) \stackrel{\text { def }}{=}\left|I \star \text{DoG}_{\sigma}\right|^{+}
\end{equation}
where $\left| \cdot\right|^{+}$ is the halfwave rectification function to make all negative value to zero.

\item[(ii)] \textit{Configuring B-COSFIRE filter}: After computing the DoG filter response, the B-COSFIRE filter is configured by considering a number of concentric circles around the center of Gaussian function. Each concentric circle can defined by three parameters $(\sigma_i,\rho_i,\phi_i)$ for $i=1,\ldots,k$ where $k$ is the number of concentric circle. These parameters can be tuned using a set of training images. In this work, we have found these values empirically.
\item[(iii)] \textit{Blurring and Shifting Responses}: In this step, the DoG responses are combined after blurring and shifting the responses using Equation~\ref{eq_blurshift}.
\begin{equation}
s_{\sigma_{i}, \rho_{i}, \phi_{i}}(x, y)=\max _{x^{\prime}, y^{\prime}}\left\{c_{\sigma_{i}}\left(x-\Delta x_{i}-x^{\prime}, y-\Delta y_{i}-y^{\prime}\right) G_{\sigma^{\prime}}\left(x^{\prime}, y^{\prime}\right)\right\}
\label{eq_blurshift}
\end{equation}
where $G_{\sigma^{\prime}}\left(x^{\prime}, y^{\prime}\right)$ is the Gaussian kernel having standard deviation as the linear function of $\rho_i$ ($\sigma'=\sigma'_0+\alpha \rho_i$). Thereafter, each blurred DoG filter responses are displaced by an amount equal to 
$\rho_{i}$ in the direction opposite to $\psi_{i}$. This ensures that DoG filters have common center that is the support center of the B-COSFIRE filter. 
\item[(iv)] \textit{Response of B-COSFIRE filters}: The final response of the B-COSFIRE filter is computed by taking weighted geometric mean of all the blurred-shifted responses obtained in previous step. The geometric mean operation over set $|S|$ is defined mathematically as
\begin{equation}
r_{S}(x, y) \stackrel{\text { def }}{=}\left|\left(\prod_{i=1}^{|S|}\left(s_{\sigma_{i}, \rho_{i}, \phi_{i}}(x, y)\right)^{\omega_{i}}\right)^{1 / \sum_{i=1}^{|S|} \omega_{i}}\right|_{t}
\end{equation}
\noindent where,
\[\omega_{i}=e^{-\frac{\rho_{i}}{2 \hat{\sigma}^{2}}}~~~~~\text{and}~~~~~\hat{\sigma}=\frac{1}{3} \max _{i \in\{1 \ldots .|S|\}}\left\{\rho_{i}\right\}\]
where the subscript $t$ for $\left|\cdot\right|$ denote the thresholding of the response by an amount of $t$ ($0\leq t\leq 1$) of the maximum responses.
\end{itemize}    
 
\par To detect all blood vessels having different orientation, the rotated versions of the response of B-COSFIRE filters are combined using equation
~\ref{Eq_comb}
\begin{equation}
\hat{r}_{S}(x, y) \stackrel{\text { def }}{=} \max _{\psi \in \Psi}\left\{r_{R_{\psi}(S)}(x, y)\right\}
    \label{Eq_comb}
\end{equation}
where $r_{R_\psi}(S)$ correspond to rotated responses of the B-COSFIRE filter at an angle $\psi$, i.e., for parameter set $R_{\psi}(S)=\{(\sigma_i,\rho_i,\phi_i+\psi~|~\forall (\sigma_i,\rho_i,\phi_i)\in S\}$.

\subsection{Post-processing}
\par To make the blood vessel extraction process automated, an adaptive thresholding technique is applied to binarize the feature image obtained through B-COSFIRE filters. In the present work, we adapt the method proposed by Otsu \textit{et al.}~\cite{zhang2008image} To reduce the false positive, cluster of less than 11 foreground pixels are eliminated from the binary image using the connected components labelling method.


\section{Results and Discussion}
\subsection{Data sets}
For quantitative evaluation of the proposed approach, the publicly available DRIVE data set is used. This data set is very popular and widely used in the community due to the availability of its gold standard ground truth (GT) images prepared by different observers. It helps to establish a relative study and preliminary evaluation of retinal vessel segmentation algorithms.
20-20 of the 40 colour images of DRIVE dataset were used for training and testing. 
Along with each image, two gold standard ground truth and one field-of-view (FoV) mask are provided in the data set.

\subsection{Experimental results}
Different evaluation metrics are used to examine the overall performance of the proposed approach and to compare with competitive algorithms, in terms of accuracy ($Acc$), sensitivity ($Sen$), area under the ROC (AUC), specificity ($Spe$),  Kappa Agreement ($\kappa$), and Matthews correlation coefficient ($MCC$)~\cite{chicco2020the}. The pixels inside the FoV are used to compute these metrics. 
 For the mathematical form of the evaluation metrics, one may refer~\cite{Kumar2019}. 
Among these, $MCC$, gives more reliable result because it consider all the four measures of the confusion matrix (true positive, false positive, true negative, and false negative). It considers proportionally both to the number of blood vessel pixels and the background pixels in the image (imbalanced class labeled data set as in retinal fundus images).
\par
\begin{figure}
    \centering
    \subfigure[]{\includegraphics[trim={1cm 5cm 12cm 9cm},clip,scale=0.25]{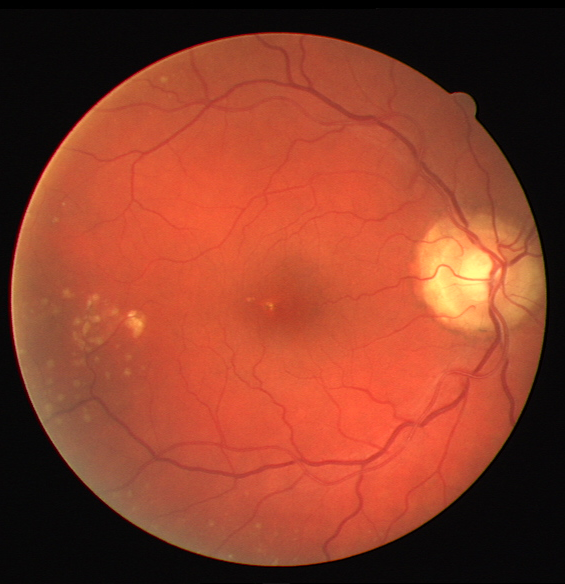}~~~\includegraphics[trim={1cm 5cm 12cm 9cm},clip,scale=0.25]{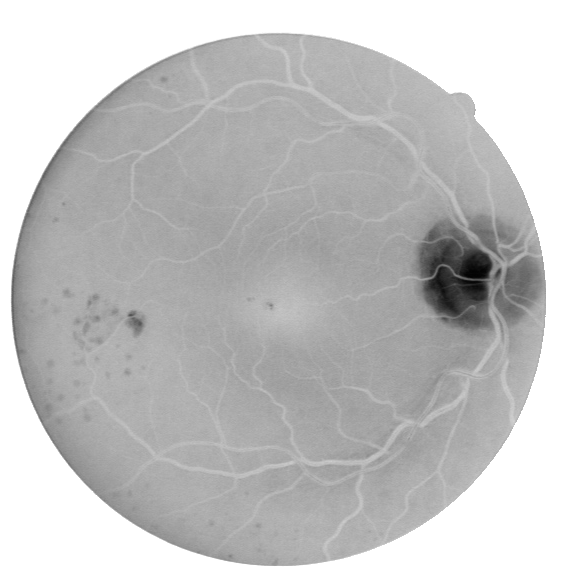}~~~\includegraphics[trim={1cm 5cm 12cm 9cm},clip,scale=0.25]{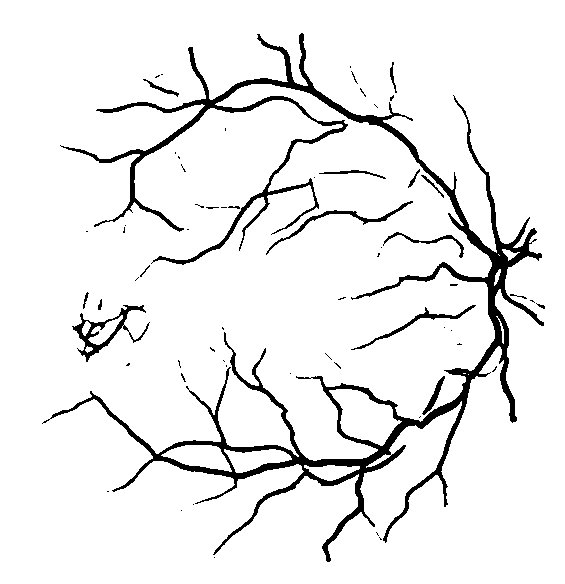}~~~\includegraphics[trim={1cm 5cm 12cm 9cm},clip,scale=0.25]{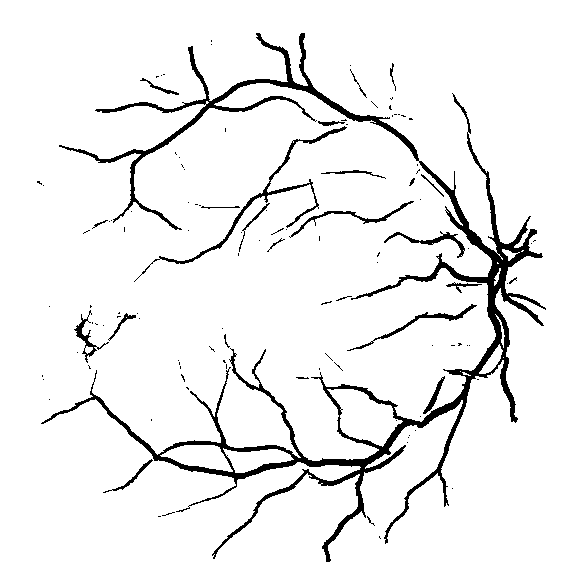}}\\
 \subfigure[]{\includegraphics[trim={1cm 8cm 12cm 6cm},clip,scale=0.25]{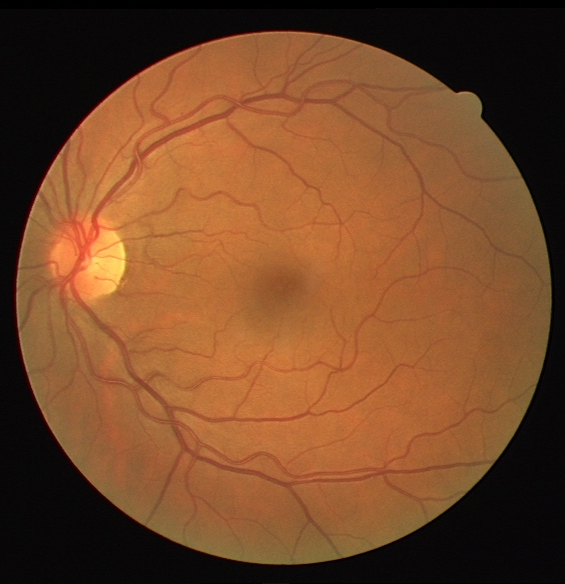}~~~\includegraphics[trim={1cm 8cm 12cm 6cm},clip,scale=0.25]{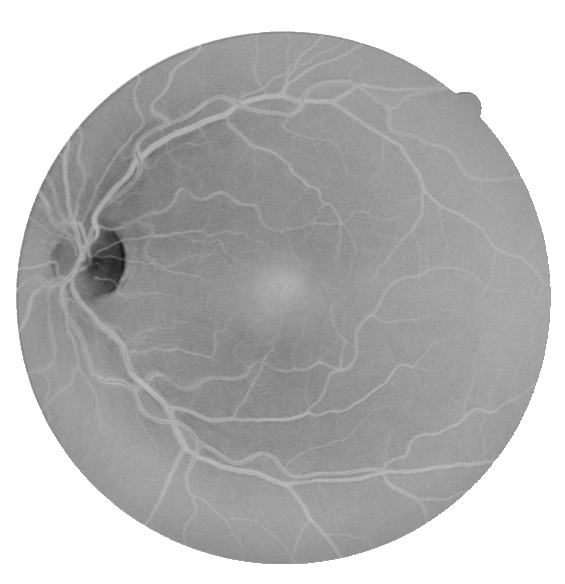}~~~\includegraphics[trim={1cm 8cm 12cm 6cm},clip,scale=0.25]{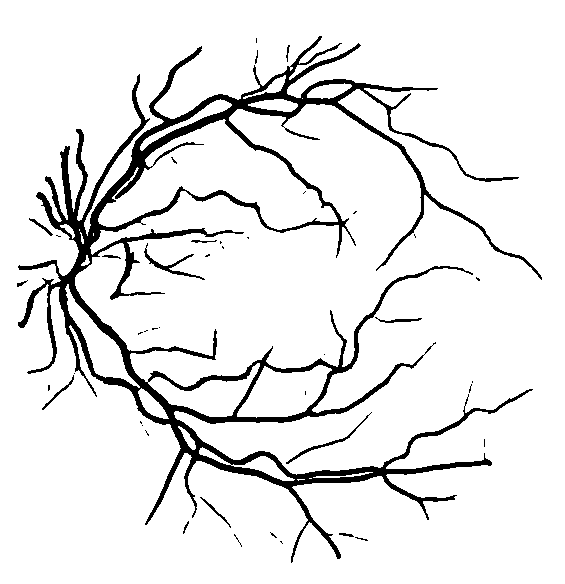}~~~\includegraphics[trim={1cm 8cm 12cm 6cm},clip,scale=0.25]{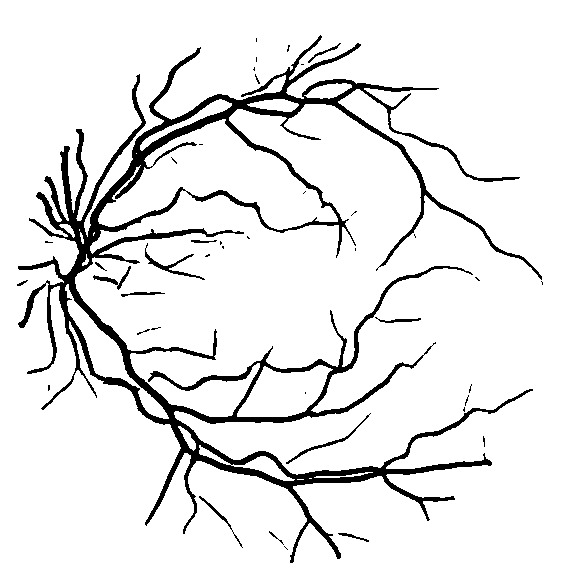}}
    \caption{Enlarged pathological region and optic disc from original color image (first column), Inverted green channel (second column), Segmented image without using Top-Hat based preprocessing technique (third column), and Segmented image using Top-Hat based preprocessing technique (fourth column)}
    \label{fig_enlarged}
\end{figure}

 In DRIVE test data set, image 3 and image 8 are two pathological fundus images. Figure~\ref{fig_enlarged} shows the enlarged pathological region (where soft and hard exudates are present) and optic disc of retinal image and their corresponding segmented results with and without Top-hat transformation in the preprocessing step, indicating higher efficacy of the proposed approach to suppress the structures other than the blood vessels.

\par We compared our results with competitive algorithms in the domain and a summary of the same is presented in Table~\ref{tab_comp}. We observe that the presented approach is better than many other reported algorithms in the literature. Specifically, in comparison with the B-COSFIRE filter proposed by Azzopardi \textit{et al.}, the $MCC$ value shows noticeable improvement. It tells that there is a proper balance between sensitivity and specificity.

\begin{table}[!h]
\vspace{-0.4cm}
\caption{A comparison of the present work with competitive algorithms in the domain on the DRIVE test data set}
\label{tab_comp}
\begin{center}
\begin{scriptsize}
\vspace{-0.2cm}
\begin{tabular}{L{3.3cm}C{1cm}C{1cm}C{2.3cm}C{1cm}C{1cm}C{1cm}}
\toprule
\multicolumn{1}{l}{Approaches}     & $Sen$    & $Spe$   & Acc  ($\sigma$)           &   $AUC$   & $\kappa$ & MCC\\ \midrule\midrule
\multicolumn{6}{L{9cm}}{\textbf{Supervised Methods}}\\\midrule
Soares \textit{et al.}~\cite{Soares2006} & 0.7230 & 0.9762 & 0.9466~~~~~~~~(-) &  0.9614  & -    &   - \\
Staal \textit{et al.}~\cite{Staal2004}   & 0.7194 & 0.9773 & 0.9441~~~~~~~~(-) &  0.9520  & -      & - \\
Ricci \textit{et al.}~\cite{Ricci2007}   &   -    &      - & 0.9595~~~~~~~~(-) &  0.9558  & -     &  - \\
\midrule\midrule
\multicolumn{6}{L{9cm}}{\textbf{Unsupervised Methods}}\\\midrule
2nd observer                    & 0.7760 & 0.9725 & 0.9473~~~~~~~~(-) &    -     & 0.6970  &  -\\
Chaudhuri \textit{et al.}\cite{Chaudhuri1989}$^{*}$ &    -    &    -    & 0.8773 (0.0232)  &  0.7878   & 0.3357    & - \\ 
Zana \textit{et al.}~\cite{Zana2001}$^{*}$   &    -   &     -   & 0.9377 (0.0077) & 0.8984   & 0.6971 &  - \\ 
Zhang \textit{et al.}~\cite{Zhang2010} &   0.7120   &   0.9724   & 0.9382~~~~~~~~(-)          &    -     & - & -\\
Zhao \textit{et al.}~\cite{Zhao2014} &       0.7354 &    0.9789    & 0.9477~~~~~~~~(-)           & -        & -  &  - \\
Gou \textit{et al.}
\cite{Gou2018} &       0.7526 &    0.9669    & 0.9393~~~~~~~~(-)           & -        & -  &   -\\ 
Kumar \textit{et al.}~\cite{Kumar2019}                & 0.7503 & 0.9717 & 0.9432 (0.0049)          & 0.9524 & 0.7374  & -
 \\ 
Azzopardi \textit{et al.}~\cite{Azzopardi2015}                 & 0.7655 & 0.9704 & 0.9442~~~~~~~~(-)          & 0.9614 & (-) & 0.7475
\\ 
Presented approach                 & 0.7531 & 0.9796 & 0.9466 (0.0042)          & 0.9603 & 0.7439   & 0.7482
\\ 
\midrule
\multicolumn{6}{L{9cm}}{$^{*}$results are taken from ~\cite{Niemeijer2004}}\\\bottomrule
\end{tabular}
\vspace{-1cm}
\end{scriptsize}
\end{center}
\end{table}


\section{Conclusion}
An effective and efficient blood vessel extraction approach has been proposed by integrating a top-hot based preprocessing technique and a fine-tuned B-COSFIRE filter. Top-hat based preprocessing technique is much capable to handle many challenges in the preprocessing step itself. The segmented results show that our proposed approach is effective to separate the vasculature from the background having other structures such as fovea, exudates, optic disc, etc., especially in pathological fundus images. On DRIVE dataset, we achieved the $Sen = 0.7531$, $Spe=0.9796$, $Acc = 0.9466$, $AUC$ = 0.9603, $\kappa = 0.7439$, and $MCC= 0.7482$ that are notably higher than many previously reported algorithms in the domain. In terms of Accuracy and Matthews correlation coefficient (MCC), the presented work beats many state-of-art algorithms. Moreover, the submitted approach is entirely unsupervised, simple, and easily implementable.

\bibliographystyle{splncs03_unsrt}
\bibliography{mybib.bib}

\begin{thebibliography}{10}
\providecommand{\url}[1]{\texttt{#1}}
\providecommand{\urlprefix}{URL }

\bibitem{Chaudhuri1989}
Chaudhuri, S., Chatterjee, S., Katz, N., Nelson, M., Goldbaum, M.: Detection of
  blood vessels in retinal images using two-dimensional matched filters. IEEE
  Transactions on medical imaging  8(3),  263--269 (1989)

\bibitem{mookiah2021a}
{Mookiah}, M.R.K., {Hogg}, S., {MacGillivray}, T.J., {Prathiba}, V.,
  {Pradeepa}, R., {Mohan}, V., {Anjana}, R.M., {Doney}, A.S., {Palmer}, C.N.,
  {Trucco}, E.: A review of machine learning methods for retinal blood vessel
  segmentation and artery/vein classification. Medical Image Analysis  68,
  101905 (2021)

\bibitem{Soares2006}
Soares, J.V., Leandro, J.J., Cesar, R.M., Jelinek, H.F., Cree, M.J.: Retinal
  vessel segmentation using the 2-{D} {G}abor wavelet and supervised
  classification. IEEE Transactions on medical Imaging  25(9),  1214--1222
  (2006)

\bibitem{Fraz2012}
Fraz, M.M., Remagnino, P., Hoppe, A., Uyyanonvara, B., Rudnicka, A.R., Owen,
  C.G., Barman, S.A.: Blood vessel segmentation methodologies in retinal
  images--a survey. Computer methods and programs in biomedicine  108(1),
  407--433 (2012)

\bibitem{Hoover2000}
Hoover, A., Kouznetsova, V., Goldbaum, M.: Locating blood vessels in retinal
  images by piecewise threshold probing of a matched filter response. IEEE
  Transactions on Medical imaging  19(3),  203--210 (2000)

\bibitem{Zhang2010}
Zhang, B., Zhang, L., Zhang, L., Karray, F.: Retinal vessel extraction by
  matched filter with first-order derivative of {G}aussian. Computers in
  biology and medicine  40(4),  438--445 (2010)

\bibitem{8989144}
Mallick, D., Kumar, K., Agarwal, S.: Blood vessel detection using modified
  multiscale {MF-FDOG} filters for diabetic retinopathy. In: 2019 International
  Conference on Applied Machine Learning (ICAML). pp. 82--86 (2019)

\bibitem{Fu2016a}
Fu, H., Xu, Y., Lin, S., Wong, D.W.K., Liu, J.: Deepvessel: Retinal vessel
  segmentation via deep learning and conditional random field. In:
  International conference on medical image computing and computer-assisted
  intervention. pp. 132--139. Springer (2016)

\bibitem{Jin2019a}
Jin, Q., Meng, Z., Pham, T.D., Chen, Q., Wei, L., Su, R.: Dunet: A deformable
  network for retinal vessel segmentation. Knowledge-Based Systems  178,
  149--162 (2019)

\bibitem{Li2020}
Li, L., Verma, M., Nakashima, Y., Nagahara, H., Kawasaki, R.: Iternet: Retinal
  image segmentation utilizing structural redundancy in vessel networks. In:
  The IEEE Winter Conference on Applications of Computer Vision (WACV) (March
  2020)

\bibitem{Odstrcilik2013}
Odstrcilik, J., Kolar, R., Budai, A., Hornegger, J., Jan, J., Gazarek, J.,
  Kubena, T., Cernosek, P., Svoboda, O., Angelopoulou, E.: Retinal vessel
  segmentation by improved matched filtering: evaluation on a new
  high-resolution fundus image database. IET Image Processing  7(4),  373--383
  (2013)

\bibitem{Azzopardi2015}
Azzopardi, G., Strisciuglio, N., Vento, M., Petkov, N.: Trainable {COSFIRE}
  filters for vessel delineation with application to retinal images. Medical
  image analysis  19(1),  46--57 (2015)

\bibitem{Kumar2019}
Kumar, K., Samal, D., Suraj: Automated retinal vessel segmentation based on
  morphological preprocessing and 2{D}-{G}abor {W}avelets. In: Advanced
  Computing and Intelligent Engineering. pp. 411--423. Springer Singapore,
  Singapore (2020)

\bibitem{Azzopardi2013}
Azzopardi, G., Petkov, N.: Automatic detection of vascular bifurcations in
  segmented retinal images using trainable cosfire filters. Pattern Recognition
  Letters  34(8),  922--933 (2013)

\bibitem{zhang2008image}
Zhang, J., Hu, J.: Image segmentation based on 2d otsu method with histogram
  analysis. In: 2008 international conference on computer science and software
  engineering. vol.~6, pp. 105--108. IEEE (2008)

\bibitem{chicco2020the}
{Chicco}, D., {Jurman}, G.: The advantages of the matthews correlation
  coefficient (mcc) over f1 score and accuracy in binary classification
  evaluation. BMC Genomics  21(1),  1--13 (2020)

\bibitem{Staal2004}
Staal, J., Abr{\`a}moff, M.D., Niemeijer, M., Viergever, M.A., Van~Ginneken,
  B.: Ridge-based vessel segmentation in color images of the retina. IEEE
  transactions on medical imaging  23(4),  501--509 (2004)

\bibitem{Ricci2007}
Ricci, E., Perfetti, R.: Retinal blood vessel segmentation using line operators
  and support vector classification. IEEE transactions on medical imaging
  26(10),  1357--1365 (2007)

\bibitem{Zana2001}
Zana, F., Klein, J.C.: Segmentation of vessel-like patterns using mathematical
  morphology and curvature evaluation. IEEE transactions on image processing
  10(7),  1010--1019 (2001)

\bibitem{Zhao2014}
Zhao, Y.Q., Wang, X.H., Wang, X.F., Shih, F.Y.: Retinal vessels segmentation
  based on level set and region growing. Pattern Recognition  47(7),
  2437--2446 (2014)

\bibitem{Gou2018}
Gou, D., Wei, Y., Fu, H., Yan, N.: Retinal vessel extraction using dynamic
  multi-scale matched filtering and dynamic threshold processing based on
  histogram fitting. Machine Vision and Applications  29(4),  655--666 (2018)

\bibitem{Niemeijer2004}
Niemeijer, M., Staal, J., van Ginneken, B., Loog, M., Abramoff, M.D.:
  Comparative study of retinal vessel segmentation methods on a new publicly
  available database. In: Medical Imaging 2004: Image Processing. vol. 5370,
  pp. 648--657. International Society for Optics and Photonics (2004)

\end{thebibliography}
\end{document}